# Linear Orbit Parameters for the Exact Equations of Motion*

George Parzen

January 11, 1994

ABSTRACT

This paper defines the beta function and other linear orbit parameters using the exact equations of motion. The $\beta$, $\alpha$ and $\psi$ functions are redefined using the exact equations. Expressions are found for the transfer matrix and the emittance. The differential equation for $\eta = x/\beta^{1/2}$ is found. New relationships between $\alpha$, $\beta$, $\psi$ and $\nu$ are derived.

## 1. Introduction

This paper defines the beta function and the other linear orbit parameters using the exact equations of motion. The usual treatment[1] of the linear orbit parameters is based on the approximate equation of motion

$$\frac{d^2 x}{ds^2} + K(s)\, x = 0 \tag{1.1}$$

Approximations are made in obtaining Eq. (1.1) which are usually valid for large accelerators.

The exact linearized equations of motion can be written as

$$\begin{aligned}\frac{dx}{ds} &= A_{11} x + A_{12} p_x \\ \frac{dp_x}{ds} &= A_{21} x + A_{22} p_x\end{aligned} \tag{1.2}$$

acc-phys/9508001 31 Aug 1995

$x$ and $p_x$ are the canonical coordinates in in a curvilinear coordinate system based on a reference orbit and the $A_{ij}(s)$ are periodic in $s$ with period $L$. The approximate Eq. (1.1) assumes that $A_{11} = A_{22} = 0$, $A_{12} = 1$ and $A_{21} = -K(s)$. The exact values of the $A_{ij}$ are given in section 2.

A treatment of the linear orbit parameters based on the exact equations, Eqs. (1.2), rather than the approximate Eq. (1.1) may be desirable in the following situations:

1. Symplectic long term tracking using a procedure where the magnets are replaced by a sequence of point magnets and drift spaces. For the tracking to be symplectic, one has to use the solutions of the exact equations of motion. The linearized equations of motion then have the form of Eq. (1.2).
2. Small accelerators where the approximation made in deriving Eq. (1.1) may not be valid.

Many of the results found using the approximate equations carry over for the exact equations. A few of the changed results are the following:

$$\alpha = \frac{1}{A_{12}} \left( -\frac{1}{2} \frac{d\beta}{ds} + A_{11}\beta \right)$$
$$\psi = \int_0^s A_{12} \frac{ds}{\beta} \qquad (1.3)$$
$$\nu = \frac{1}{2\pi} \int_0^C ds \frac{A_{12}}{\beta}$$

where $C$ is the circumference of the accelerator.

Some unchanged results are

$$\epsilon = \gamma x^2 + 2\alpha\gamma p + \beta p^2$$
$$M(s+L, s) = \begin{pmatrix} \cos\mu + \alpha\sin\mu & \beta\sin\mu \\ -\gamma\sin\mu & \cos\mu - \alpha\sin\mu \end{pmatrix} \qquad (1.4)$$
$$\gamma = (1 + \alpha^2)/\beta$$
$$\mu = \psi(s+L) - \psi(s)$$

$L$ is the period of the $A_{ij}(s)$. More detailed results are given below. Because of Eqs. (1.4), the usual procedure used in tracking programs to compute $\nu$, $\beta$, $\alpha$, $\gamma$ and $\psi$, from the transfer matrix is still valid.

## 2. Equations of Motion

The exact equations of motion can be written[1,2] as

$$\frac{dx}{ds} = \frac{1 + x/\rho}{q_s} q_x$$

$$\frac{dq_x}{ds} = \frac{q_s}{\rho} + \frac{1}{B\rho}\left[(1 + x/\rho) B_y - \frac{q_y}{q_s}(1 + x/\rho) B_s\right]$$

$$\frac{dy}{ds} = \frac{1 + x/\rho}{q_s} q_y \qquad (2.1)$$

$$\frac{dq_y}{ds} = \frac{1}{B\rho}\left[\frac{q_x}{q_s}(1 + x/\rho) B_s - (1 + x/\rho) B_x\right]$$

$$q_s = \left(1 - q_x^2 - q_y^2\right)^{1/2}, \qquad B\rho = pc/e$$

In Eq. (2.1), $x$, $s$, $y$ are the particle coordinates in a curvilinear coordinate system based on a reference orbit with the radius curvature $\rho(s)$. $p_x$, $p_s$, $p_y$ are the components of the momentum and

$$q_x = p_x/p, \qquad q_y = p_y/p, \qquad q_s = p_s/p \qquad (2.2)$$

For large accelerators, $q_x \simeq x'$ and $q_y \simeq y'$.

To find the linearized equations for the betatron oscillations, one expands Eq. (2.1) around the closed orbit for a particular momentum, $p$. This gives the set of linear equations

$$\frac{d\bar{x}_i}{ds} = \sum_j A_{ij} \bar{x}_j \qquad (2.3a)$$

where $\bar{x}_i$ are the 4 coordinates relative to the closed orbit. The $A_{ij}$ are given by

$$A_{ij} = \frac{\partial}{\partial x_j} f_i \qquad i,j = 1,4 \qquad (2.3b)$$

evaluated on the closed orbit and Eqs. (2.1) have been written as

$$\frac{dx_i}{ds} = f_i \qquad i = 1,4 \qquad (2.3c)$$

For the case where $B_s = 0$, $q_y = 0$, $y = 0$ on the closed orbit, the $A_{ij}$ are given by

$$A_{ij} = \begin{bmatrix} \dfrac{1}{\rho}\dfrac{q_x}{q_s} & \dfrac{(1 + x/\rho)}{q_s^3} & 0 & 0 \\ \dfrac{1}{B\rho}\left[\dfrac{B_y}{\rho} + (1 + x/\rho)\dfrac{\partial B_y}{\partial x}\right] & -\dfrac{1}{\rho}\dfrac{q_x}{q_s} & 0 & 0 \\ 0 & 0 & 0 & \dfrac{(1 + x/\rho)(1 - q_x^2)}{q_s^3} \\ 0 & 0 & -\dfrac{1}{B\rho}(1 + x/\rho)\dfrac{\partial B_x}{\partial y} & 0 \end{bmatrix}$$

$$(2.3d)$$

For large accelerators where $q_x \ll 1$, $q_y \ll 1$, $x/\rho \ll 1$ one has

$$A_{11} = A_{22} = 0, \quad A_{12} = 1$$
$$A_{34} = 1.$$
(2.4)

For the exact equations, $A_{11}$ and $A_{22}$ are not zero, and one does not have $A_{12} = 1$, $A_{34} = 1$. In particular,

$$A_{11} = -A_{22} = (1/\rho)(q_x/q_s)$$
$$A_{12} = (1 + x/\rho)/q_s^3$$
$$A_{34} = (1 + x/\rho)\left(1 - q_x^2\right)/q_s^3$$
(2.5)

## 3. Eigenfunctions of the Exact Linear Equations of Motion and the Linear Orbit Parameters

The problem now is, given the exact linear equations of motion, Eq. (2.3), how does one define the linear orbit parameters $\beta$, $\alpha$, $\gamma$, $\nu$ and the emittance $\epsilon$, and what are the relationships that hold between them. To do this, one has to repeat the well known treatment of the linear orbit parameters, and see where the definitions and relationships change for the exact equations. The treatment given below is believed to reduce the amount of algebraic manipulation required, and makes few assumptions about the $A_{ij}$ coefficients in the linear equations.

For the $x$ motion, the linear equations are written as

$$\frac{dx}{ds} = A_{11}x + A_{12}p_x$$
$$\frac{dp_x}{ds} = A_{21}p_x + A_{22}x$$
(3.1)

The $A_{ij}$ are given by Eqs. (2.3).

The transfer matrix $M(s, s_0)$ obeys

$$x = M(s, s_0)x_0$$
$$x = \begin{pmatrix} x \\ p_x \end{pmatrix}$$
$$\frac{d}{ds}M = AM$$
(3.2)

One may note that the symbol $x$ is used in 2 different ways. The meaning of $x$ should be clear from the context. The matrix $M$ is symplectic as the equations of motion are derived from a hamiltonian.[1,2] Thus

$$M\overline{M} = I$$
$$\overline{M} = \tilde{S}\tilde{M}\,S \qquad (3.3)$$
$$S = \begin{pmatrix} 0 & 1 \\ -1 & 0 \end{pmatrix}, \quad I = \begin{pmatrix} 1 & 0 \\ 0 & 1 \end{pmatrix}$$

$\tilde{S}$ is the transverse of $S$. Also $|M| = 1$; $|M|$ is the determinant of $M$.

The one period transfer matrix is defined by

$$\widehat{M}(s) = M(s+L, s) \qquad (3.4a)$$

where $L$ is the period of the $A_{iy}$ in Eq. (3.1). One can show that $\widehat{M}(s)$ and $\widehat{M}(s_0)$ are related by

$$\widehat{M}(s) = M(s, s_0)\,\widehat{M}(s_0)\,M(s_0, s) \qquad (3.4b)$$

The eigenfunctions and eigenvalues of $\widehat{M}(s)$ are defined by

$$\widehat{M}(s)\,x = \lambda x,$$
$$|\widehat{M} - \lambda I| = 0, \qquad (3.5)$$
$$\lambda^2 - (m_{11} + m_{22})\lambda + 1 = 0$$

where $m_{ij}$ are the elements of $\widehat{M}$, and using $|\widehat{M}| = 1$.

Eqs. (3.5) shows that the two eigenvalues $\lambda_1$, $\lambda_2$ obey

$$\lambda_1 \lambda_2 = 1, \qquad (3.6a)$$

and for stable motion, $|\lambda| = 1$ and $\lambda_2 = \lambda_1^*$, and we can write

$$\lambda_1 = \exp(i\mu) \qquad (3.6b)$$

Given the eigenfunction at $s_0$, $x_1(s_0)$ one can find the eigenfunction at any other point $s$ using

$$x_1(s) = M(s, s_0)\,x_1(s_0), \qquad (3.7a)$$

and $x_1(s)$ has the same eigenvalue $\lambda_1$. This follows from Eq. (3.5), using Eq. (3.4b) to relate $\widehat{M}(s)$ and $\widehat{M}(s_0)$. Also $x_1(s)$ obeys the linear equations of motion,

$$\frac{d}{ds}x_1 = Ax_1, \tag{3.7b}$$

which follows from Eq. (3.7a) and Eq. (3.2). One can show that

$$x_1(s)/\lambda_1^{s/L} = f_1(s) \tag{3.8a}$$

where $f_1(s+L) = f_1(s)$. This follows from

$$f_1(s+L) = x_1(s+L)/\lambda_1^{s/L+1}$$
$$= \widehat{M}(s)x_1(s)/\lambda_1^{s/L+1} = x_1(s)/\lambda_1^{s/L}$$

Thus, one can write

$$x_1(s) = \exp(i\mu s/L)f_1(s)$$
$$f_1(s+L) = f_1(s) \tag{3.8b}$$

Eq. (3.8b) can be rewritten as

$$x_1(s) = \beta(s)^{1/2}\exp(i\psi)$$
$$\psi(s) = \mu s/L + g_1(s) \tag{3.9}$$
$$g_1(s+L) = g_1(s), \quad \beta(s+L) = \beta(s)$$

Eq. (3.9) defines the beta functions, $\beta(s)$, except for a normalization multiplyer, for the eigenfunction $x_1(s)$. The normalization multiplyer will be defined below. It will be shown first that $\psi$ and $\beta$ are related. To find this relation, one uses the Lagrange invariant[1]

$$W = \tilde{x}_2 S x_1 \tag{3.10}$$

where $x_1$, $x_2$ are two solutions of the equations of motion. Eq. (3.10) corresponds to the Wronskian in the treatment of the approximate equations of motion. For $x_1$ and $x_2$, we use the two eigenfunctions $x_1$ and $x_2 = x_1^*$.

$$x_1 = \begin{pmatrix} x_1 \\ p_{x1} \end{pmatrix} \tag{3.11}$$

For $x_1$ one uses Eq. (3.9) and for $p_{x1}$ one finds from the equations of motion

$$p_{x1} = \frac{1}{A_{12}}\left(\frac{dx_1}{ds} - A_{11}x\right) \tag{3.12}$$

$$W = x_2 p_{x1} - p_{x2} x_1$$
$$W = \left[x_2\frac{dx_1}{ds} - \frac{x_1 dx_2}{ds}\right]\frac{1}{A_{12}} \tag{3.13}$$
$$W = \frac{2i}{A_{12}}\beta\frac{d\psi}{ds}$$

The beta function $\beta$ is normalized by normalizing the eigenfunctions so that

$$W = \tilde{x}_1^* S x_1 = 2i \qquad (3.14)$$

which gives

$$\frac{d\psi}{ds} = \frac{A_{12}}{\beta} \qquad (3.15)$$

Eq. (3.15) replaces the familiar result $d\psi/ds = 1/\beta$ which is obtained when $A_{12} = 1$. From Eq. (3.15) one can find a result for the tune. Using $2\pi\nu = \psi(C) - \psi(0)$ where $C$ is the circumference of the accelerator, one finds

$$\nu = \frac{1}{2\pi} \int_0^C ds \frac{A_{12}}{\beta} \qquad (3.16)$$

¿From Eq. (3.12) we now find for $p_{x1}$,

$$p_{x1} = \frac{1}{\beta^{1/2}} (i - \alpha) \exp(i\psi) \qquad (3.17\text{a})$$

$$\alpha = \frac{1}{A_{12}} \left( -\frac{1}{2} \frac{d\beta}{ds} + A_{11}\beta \right) \qquad (3.17\text{b})$$

Eqs. (3.17) provide the new definition for the $\alpha$ parameter, which replaces the familiar result $\alpha = \frac{1}{2} d\beta/ds$. At this point the definition of $\alpha$ may seem arbitrary. It will be seen to be the convenient definition of $\alpha$ when the emittance and transfer matrix are considered below.

The eigenfunctions can now be written as, using Eq. (3.9) and Eq. (3.17),

$$x_1 = \begin{bmatrix} \beta^{\frac{1}{2}} \\ \beta^{-\frac{1}{2}} (-\alpha + i) \end{bmatrix} \exp(i\psi) \qquad (3.18)$$

$$x_2 = x_1^*$$

## 4. The Transfer Matrix, the Emittance and the Linear Orbit Parameters

The particle motion can be written as a linear combination of the eigenfunction given by Eq. (3.18)

$$x = ax_1 + c.c.$$

$$x = 2|a|\beta^{1/2}\cos(\psi + \delta)$$
$$p_x = -\frac{\alpha x}{\beta} - \frac{2|a|\sin(\psi + \delta)}{\beta^{1/2}} \quad (4.1)$$
$$a = |a|\exp(i\delta)$$

Eqs. (4.1) suggest the new variables $\eta$, $p_\eta$

$$\begin{pmatrix} \eta \\ p_\eta \end{pmatrix} = G \begin{pmatrix} x \\ p_x \end{pmatrix}$$
$$G = \begin{bmatrix} \beta^{-\frac{1}{2}} & 0 \\ \alpha/\beta^{\frac{1}{2}} & \beta^{\frac{1}{2}} \end{bmatrix} \quad (4.2)$$

for then

$$\eta = 2|a|\cos(\psi + \delta)$$
$$p_n = -2|a|\sin(\psi + \delta) \quad (4.3)$$

one obtains the emittance invariant from Eq. (4.3)

$$\eta^2 + p_\eta^2 = \epsilon$$
$$2|a| = \epsilon^{1/2} \quad (4.4)$$
$$\eta = \epsilon^{1/2}\cos(\psi + \delta), \quad p_n = -\epsilon^{1/2}\sin(\psi + \delta)$$

Replacing $\eta$, $p_\eta$, by $x$, $p_x$ using Eq. (4.2), one gets

$$\epsilon = \gamma x^2 + 2\alpha x p_x + \beta p_x^2$$
$$\gamma = (1 + \alpha^2)/\beta \quad (4.5)$$

It will be shown that $\epsilon$ is the phase space area, divided by $\pi$, inside the ellipse defined by Eq. (4.5).

Since $|G| = 1$, $G$ is a symplectic matrix and $\eta$, $p_\eta$ are symplected variables with the transfer matrix $U(s, s_0)$ and

$$\begin{pmatrix} \eta \\ p_\eta \end{pmatrix} = U(s, s_0) \begin{pmatrix} \eta_0 \\ p_{\eta 0} \end{pmatrix} \quad (4.6)$$
$$U = \overline{G}(s) M(s, s_0) G(s_0)$$

$$\overline{G} = G^{-1} = \begin{pmatrix} \beta^{1/2} & 0 \\ -\alpha/\beta^{1/2} & \beta^{-1/2} \end{pmatrix}$$

$\widehat{U} = U(s+L, s)$ has the same eigenvalues as $\widehat{M}$, and the eigenfunction of $\widehat{U}$, $\eta_1$ and $\eta_2 = \eta_1^*$ are related to the eigenfunctions of $\widehat{M}$, $x_1$ and $x_2 = x_1^*$, by

$$\begin{aligned} \eta_1 &= G x_1 \\ \eta_1 &= \begin{bmatrix} 1 \\ i \end{bmatrix} e^{i\psi} \end{aligned} \quad (4.7)$$

where Eq. (3.18) was used for $x_1$.

One sees from Eqs. (4.3) that $p_\eta = d\eta/d\psi$ and thus Eq. (4.3) can be rewritten as

$$\begin{aligned} \eta &= \eta_0 \cos(\psi - \psi_0) + p_{\eta 0} \sin(\psi - \psi_0) \\ p_\eta &= -\eta_0 \sin(\psi - \psi_0) + p_{\eta 0} \cos(\psi - \psi_0) \end{aligned} \quad (4.8)$$

Eq. (4.8) gives a result for $U(s, s_0)$

$$U(s, s_0) = \begin{bmatrix} \cos(\psi - \psi_0) & \sin(\psi - \psi_0) \\ -\sin(\psi - \psi_0) & \cos(\psi - \psi_0) \end{bmatrix}. \quad (4.9)$$

One can then find $M(s, s_0)$ using $M = G(s) U \overline{G}(s_0)$,

$$M(s, s_0) = \begin{bmatrix} \left(\frac{\beta}{\beta_0}\right)^{1/2} [\cos(\psi - \psi_0) + \alpha_0 \sin(\psi - \psi_0)] & (\beta \beta_0)^{1/2} \sin(\psi - \psi_0) \\ -(1 + \alpha \alpha_0) \frac{\sin(\psi - \psi_0)}{(\beta \beta_0)^{1/2}} - (\alpha - \alpha_0) \frac{\cos(\psi - \psi_0)}{(\beta \beta_0)^{1/2}} & \left(\frac{\beta_0}{\beta}\right)^{1/2} [\cos(\psi - \psi_0) - \alpha \sin(\psi - \psi_0)] \end{bmatrix} \quad (4.10)$$

One then finds

$$\begin{aligned} \widehat{M} = M(s+L, s) &= \begin{bmatrix} \cos\mu + \alpha \sin\mu & \beta \sin\mu \\ -\gamma \sin\mu & \cos\mu - \alpha \sin\mu \end{bmatrix} \\ \mu &= \psi(s+L) - \psi(s) \end{aligned} \quad (4.11)$$

The results for $M$ and $\widehat{M}$ are unchanged from those found for the approximate equations of motion. These results and the result for the emittance Eq. (4.5) justify the new definition for $\alpha$, Eq. (3.17).

The connection between the emittance, defined by Eq. (4.5), and the phase space area inside the ellipse area defined by Eq. (4.5), is given by

$$\int dx\, dp_x = \int d\eta\, dp_\eta = \pi \epsilon \quad (4.12)$$

where we have used $|G| = 1$ and Eq. (4.4) for the ellipse in $\eta$, $p_\eta$ space.

The relationship between the parameters $\beta$, $\alpha$, $\gamma$ at $s$ and $\beta$, $\alpha$, $\gamma$ at $s_0$ is unchanged. To see this, write

$$\widehat{M} = I \cos u + J \sin u$$
$$J = \begin{pmatrix} \alpha & \beta \\ -\gamma & -\alpha \end{pmatrix} \tag{4.13}$$

Using Eq. (3.4b) that connects $\widehat{M}(s)$ and $\widehat{M}(s_0)$ one finds

$$J = M(s, s_0) J(s_0) M^{-1}(s, s_0) \tag{4.14}$$

Eq. (4.13) gives the desired result

$$\begin{bmatrix} \beta \\ \alpha \\ \gamma \end{bmatrix} = \begin{bmatrix} m_{11}^2 & -2m_1 m_2 & m_{12}^2 \\ -m_{21} m_{11} & 1 + 2m_{12} m_{21} & -m_{12} m_{22} \\ m_{21}^2 & -2m_{22} m_{21} & m_{22}^2 \end{bmatrix} \begin{bmatrix} \beta_0 \\ \alpha_0 \\ \gamma_0 \end{bmatrix} \tag{4.15}$$

## 5. Differential Equations for the Linear Orbit Parameter

This section finds differential equations for $\beta$, $\eta$ and $\beta$, $\alpha$, $\gamma$.

### 5.1. Second Order Differential Equation for $x$

¿From the first order differential equation for $x$, $p_x$, Eq. (3.1), one can eliminate $p_x$ to find a second order equation for $x$. ¿From Eq. (3.1)

$$p = \frac{1}{A_{12}} \left( \frac{dx}{ds} - A_{11} x \right)$$
$$\frac{dp}{ds} = \frac{d}{ds} \left[ \frac{1}{A_{12}} \left( \frac{dx}{ds} - A_{11} x \right) \right] = A_{21} x + \frac{A_{22}}{A_{12}} \left( \frac{ds}{ds} - A_{11} x \right) \tag{5.1}$$
$$\frac{d}{ds} \left( \frac{1}{A_{12}} \frac{ds}{ds} \right) + x \left( -A_{21} - \frac{d}{ds} \left( \frac{A_{11}}{A_{12}} \right) - \frac{A_{11}^2}{A_{12}} \right) = 0$$

It has been assumed that $A_{11} = -A_{22}$.

## 5.2. Differential Equation for $\beta$

To find a differential equation for $\beta$, into Eq. (5.1) for $x$ put the eigenfunction

$$x = b \exp(i\psi)$$
$$b = \beta^{1/2}$$
(5.2)

We find then,

$$\frac{dx}{ds} = \left(\frac{db}{db} + \frac{iA_{12}}{b}\right) \exp(i\psi)$$
(5.3)

using $d\psi/ds = A_{12}/b^2$, Eq. (3.15)

$$\frac{d}{ds}\left(\frac{1}{A_{12}}\frac{ds}{ds}\right) = \left[\frac{d}{ds}\left(\frac{1}{A_{12}}\frac{db}{ds}\right) - \frac{i}{b^2}\frac{db}{ds}\right] \exp(i\psi)$$
$$+ \left[\frac{db}{ds} + \frac{iA_{12}}{b}\right]\frac{1}{A_{12}}\frac{iA_{12}}{b^2} \exp(i\psi)$$
$$= \left[\frac{d}{ds}\left(\frac{1}{A_{12}}\frac{db}{ds}\right) - \frac{A_{12}}{b^3}\right] \exp(i\psi)$$
(5.4)

Putting Eq. (5.4) into Eq. (5.1), one gets

$$\frac{d}{ds}\left(\frac{1}{A_{12}}\frac{db}{ds}\right) - \frac{A_{12}}{b^3} + b\left(-A_{21} - \frac{d}{ds}\left(\frac{A_{11}}{A_{12}}\right) - \frac{A_{11}^2}{A_{12}}\right) = 0$$
(5.5)

Eq. (5.5) is a second order differential equation for $b = \beta^{1/2}$. It can be compared to the result found when $A_{12} = 1$ and $A_{11} = 0$,

$$\frac{d^2 b}{ds^2} - \frac{A_{12}}{b^3} = 0$$
(5.6)

## 5.3. Differential Equation for $\eta$

$\eta$ and $x$ are related by Eq. (4.2) which can be written as

$$x = b\,\eta, \qquad b = \beta^{1/2}$$
(5.7)

In the differential equation for $\eta$ the independent variable is $\psi$ or $\theta$ which are related to $s$ by

$$d\psi = A_{12}\frac{ds}{\beta}$$
$$d\theta = A_{12}\frac{ds}{\nu\beta}$$
(5.8)

We find $dx/ds$ and $d(A_{12}^{-1}dx/ds)/ds$ which are then substituted into Eq. (5.1) to get the equation for $\eta$, using Eq. (5.5) to eliminate derivatives of $b$.

$$\frac{dx}{ds} = \frac{db}{ds}\eta + \frac{d\eta}{d\psi}\frac{A_{12}}{b}$$

$$\frac{d}{ds}\left(\frac{1}{A_{12}}\frac{dx}{ds}\right) = \frac{d}{ds}\left(\frac{1}{A_{12}}\frac{db}{ds}\right)\eta + \frac{1}{A_{12}}\frac{db}{ds}\frac{d\eta}{d\psi}\frac{A_{12}}{b^2}$$

$$+ \frac{d^2\eta}{d\psi^2}\frac{A_{12}}{b^3} - \frac{d\eta}{d\psi}\frac{1}{b^2}\frac{db}{ds}$$

$$= \left\{\frac{A_{12}}{b^3} + b\left[A_{21} + \frac{d}{ds}\left(\frac{A_{11}}{A_{12}}\right) + \frac{A_{11}^2}{A_{12}}\right]\right\}\eta \quad (5.9)$$

$$+ \frac{d^2\eta}{d\psi^2}\frac{A_{12}}{b^3}$$

$$= b\eta\left\{A_{21} + \frac{d}{ds}\left(\frac{A_{11}}{A_{12}}\right) + \frac{A_{11}^2}{A_{12}}\right\}$$

Thus the equation for $\eta$ is

$$\frac{d^2\eta}{d\psi^2} + \eta = 0$$

$$\frac{d^2\eta}{d\theta^2} + \nu^2\eta = 0 \quad (5.10)$$

The differential equation for $\eta$ is unchanged.

## 5.4. Differential Equations for $\beta$, $\alpha$, $\gamma$

The differential equations for $\beta$, $\alpha$, $\gamma$, $\gamma = (1+\alpha^2)/\beta$ can be found starting from Eq. (4.13) and (4.14) which we write as

$$\widehat{M} = I\cos\mu + \sin\mu \quad (5.11a)$$

$$J = \begin{pmatrix} \alpha & \beta \\ -\gamma & -\alpha \end{pmatrix} \quad (5.11b)$$

$$J(s) = M(s,s_0)J(s_0)M(s_0,s)$$

We note that

$$\frac{d}{ds}M(s,s_0) = A(s)M(s,s_0)$$

$$\frac{d}{ds}M(s_0,s) = -M(s_0,s)A(s) \quad (5.12)$$

The last equation follows from $M(s,s_0)M(s_0,s) = I$. Thus we find that

$$\frac{dJ}{ds} = AJ - JA \quad (5.13)$$

Replacing $J$ using Eq. (5.11b) in Eq. (5.13) we find

$$\frac{d\beta}{ds} = (A_{11} - A_{22})\beta - 2A_{12}\alpha$$
$$\frac{d\alpha}{ds} = -A_{21}\beta - A_{12}\gamma \qquad (5.14)$$
$$\frac{d\gamma}{ds} = -2A_{21}\alpha - (A_{11} - A_{22})\gamma$$

## 6. Pertrubation Theory Using the Differential Equation for $\eta$

The equation for $\eta$, Eq. (5.10) is often used as a starting point in finding the effects of a perturbing field. The particle coordinates are measured relative to a reference orbit which is the particle motion in a known magnetic field with components $B_i$. The exact equations of motion can then be written as

$$\frac{dx_i}{ds} = \sum_j A_{ij} x_j + f_i \qquad i = 1, 4, \quad j = 1, 4 \qquad (6.1a)$$

where the $f_i$ includes all the terms not included in $\sum A_{ij} x_j$. These include terms due to fields not included in the reference field $B_i$, which may be referred as $\Delta B_i$, and nonlinear terms due to the terms in the exact equations of motion that do not depend on $B_i$.

One can see from the exact equations of motion, Eqs. (2.1) that the contributions to $f_i$ which depend explicitly on $\Delta B_i$, when $\Delta B_s = 0$, are given by

$$f_2 = \frac{1}{B\rho}(1 + x/\rho)\Delta B_y$$
$$f_4 = -\frac{1}{B\rho}(1 + x/\rho)\Delta B_x \qquad (6.1b)$$

Repeating the above derivation of Eq. (5.10) for $\eta$, including the $f_i$ terms, one finds the $\eta$ equation for the $x$-motion

$$\frac{d^2\eta}{d\theta^2} + \nu_x^2 \eta = \frac{\nu_x^2 \beta_x^{3/2}}{A_{12}} f_x$$
$$f_x = f_2 + \frac{A_{11}^2}{A_{12}} f_1 + \frac{d}{ds}\left(\frac{f_1}{A_{12}}\right) \qquad (6.2)$$
$$d\theta = A_{12} \frac{ds}{\nu_x \beta_x}$$

A similar equation can be found for the $y$ motion,

$$\frac{d^2\eta}{d\theta^2} + \nu_y^2 \eta = \frac{\nu_y^2 \beta_y^{3/2}}{A_{34}} f_y$$

$$f_y = f_4 + \frac{A_{33}^2}{A_{34}} f_3 + \frac{d}{ds}\left(\frac{f_3}{A_{34}}\right)$$

(6.3)

For the case of a gradient perturbation

$$\Delta B_y = -Gx \tag{6.4}$$

where $x = 0$ on the unperturbed closed orbit, one can use Eq. (6.2) to find the change in $\nu_x$, $\Delta\nu_x$. One finds

$$\Delta\nu_x = \frac{1}{4\pi} \int ds\, \beta_x \frac{G}{B\rho} \tag{6.5}$$

This well known result for $\Delta\nu_x$ is not changed by using the exact linear equations.